\documentclass[11pt]{article}
\pdfoutput=1
\usepackage{lp,graphicx,url,wrapfig,longtable}
\usepackage{algorithm}
\usepackage{algorithmic}
\usepackage{latexsym}
\usepackage{amsmath}
\usepackage{amssymb}
\usepackage{relsize}
\usepackage{subcaption}
\usepackage{tikz}
\usetikzlibrary{arrows,positioning,fit,shapes,decorations.pathmorphing}

\newcommand{\qed}{\rule{7pt}{7pt}}

\newcounter{listCounter}

\newcommand{\ListLengths}{\setlength{\itemsep}{0ex}\setlength{\topsep}{1ex}\setlength{\partopsep}{0ex}}

\begin{document}

\title{An Experimental Evaluation\\of the Best-of-Many Christofides' Algorithm\\for the Traveling Salesman Problem\footnote{An extended abstract of this paper is scheduled to appear in the 2015 European Symposium on Algorithms (ESA 2015).}}
\author{Kyle Genova\thanks{Address: Department of Computer Science, Cornell University, Ithaca, NY
14853.  Email: {\tt kag278@cornell.edu}. Supported in part by NSF grant CCF-1115256.} \and David P.\
Williamson\thanks{Address: School of Operations Research and
Information Engineering, Cornell University, Ithaca,
 NY 14853.  Email: {\tt dpw@cs.cornell.edu}. Supported in part by NSF grant CCF-1115256.}}

\maketitle

\begin{abstract}
Recent papers on approximation algorithms for the traveling salesman problem (TSP) have given a new variant on the well-known Christofides' algorithm for the TSP, called the {\em Best-of-Many Christofides' algorithm}.  The algorithm involves sampling a spanning tree from the solution the standard LP relaxation of the TSP, subject to the condition that each edge is sampled with probability at most its value in the LP relaxation.  One then runs Christofides' algorithm on the tree by computing a minimum-cost matching on the odd-degree vertices in the tree, and shortcutting the resulting Eulerian graph to a tour.   In this paper we perform an experimental evaluation of the Best-of-Many Christofides' algorithm to see if there are empirical reasons to believe its performance is better than that of Christofides' algorithm.  Furthermore, several different sampling schemes have been proposed; we implement several different schemes to determine which ones might be the most promising for obtaining improved performance guarantees over that of Christofides' algorithm.  In our experiments, all of the implemented methods perform significantly better than the Christofides' algorithm; an algorithm that samples from a maximum entropy distribution over spanning trees seems to be particularly good, though there are others that perform almost as well.
\end{abstract}


\section{Introduction}

In the traveling salesman problem (TSP), we are given a complete, undirected graph $G=(V,E)$ as input with costs $c_e \geq 0$ for all $e \in E$, and we must find a tour through all the vertices of minimum cost.  In what follows, we will assume that the costs obey the triangle inequality; that is, $c_{(u,w)} \leq c_{(u,v)} + c_{(v,w)}$ for all $u,v,w  \in V$.  We will sometimes refer to the {\em asymmetric} traveling salesman problem (ATSP), in which the input is a complete directed graph, and possibly $c_{(u,v)} \neq c_{(v,u)}$.

In 1976, Christofides \cite{Christofides76} gave a $\frac{3}{2}$-approximation algorithm for the TSP; an $\alpha$-approximation algorithm for the TSP is one that runs in polynomial time and returns a solution of cost at most $\alpha$ times the cost of an optimal solution.  The value $\alpha$ is sometimes known as the performance guarantee of the algorithm.  Christofides' algorithm works as follows: it computes a minimum-cost spanning tree (MST) $F$ of the input graph $G$, then finds a minimum-cost perfect matching $M$ on all the odd-degree vertices of the tree $F$.  The resulting edge set $F \cup M$ is then an Eulerian subgraph of $G$: it is connected (because $F$ is connected) and has even degree at all vertices (because we added one edge incident to all odd-degree vertices of $F$).  It is well-known that an Eulerian subgraph has an Eulerian traversal that starts at any vertex of the graph, visits all of the edges of the Eulerian subgraph, then returns to the starting vertex.  By ``shortcutting'' any previously visited vertex during the traversal, we can obtain a tour that visits each vertex exactly once and has cost no greater than the cost of the edges in $F \cup M$.

No approximation algorithm with performance guarantee better than $\frac{3}{2}$ is yet known for the TSP.  However, some progress has been made in recent years for special cases and variants of the problem.  Asadpour, Goemans, Madry, Oveis Gharan, and Saberi \cite{AsadpourGMOS10} gave an $O(\log n/\log \log n)$-approximation algorithm for the ATSP (where $n=|V|$), improving on a long-standing $O(\log n)$-approximation algorithm of Frieze, Galbiati, and Maffioli \cite{FriezeGM82}.  A sequence of improvements has been obtained in the special case of the {\em graph} TSP, in which the input to the problem is an undirected, not necessarily complete graph $G$, and the cost $c_{(u,v)}$ for each $u,v \in V$ is the number of edges in the shortest $u$-$v$ path in $G$.  In this case, Oveis Gharan, Saberi, and Singh \cite{OveisGharanSS11} were able to improve slightly on the factor of $\frac{3}{2}$.  M\"omke and Svensson \cite{MomkeS11} then gave a 1.462-approximation algorithm; Mucha \cite{Mucha14} improved the analysis of the M\"omke and Svensson algorithm to obtain a $\frac{13}{9}$-approximation algorithm.  Seb\H{o} and Vygen \cite{SeboV14}, by adding some additional ideas, gave a 1.4-approximation algorithm for graph TSP.

Another line of work considers the {\em $s$-$t$ path TSP}.  In this problem, we have the same input as the TSP, plus additional vertices $s, t \in V$.  The goal is to find a minimum-cost Hamiltonian path starting at $s$ and ending at $t$ (that is, a path that starts at $s$, ends at $t$, and visits all other vertices in between).  In 1991, Hoogeveen \cite{Hoogeveen91} showed that the natural analog of Christofides' algorithm for the $s$-$t$ TSP path problem is a $\frac{5}{3}$-approximation algorithm; this remained the best known performance guarantee for over 20 years.  In 2012, An, Kleinberg, and Shmoys \cite{AnKS12} gave a $\frac{1 + \sqrt{5}}{2}$-approximation algorithm for the problem; Seb\H{o} \cite{Sebo13} improved the analysis of the An et al.\ algorithm to obtain a 1.6-approximation algorithm.  Gao \cite{Gao14} gives a nice unification of these two results.  Very recently, Vygen \cite{Vygen15} gave a slight improvement to a 1.599-approximation algorithm.

A crucial idea in several of these results is to run Christofides' algorithm, but to start with a tree that is determined by an LP relaxation of the problem at hand, rather than the minimum-cost spanning tree.  For the TSP, a well-known relaxation of the problem is as follows.
\lps & &
& \mbox{Min} & \sum_{e \in E} c_{e} x_{e} \\
& \mbox{subject to:} \\
& & & & x(\delta(v)) = 2, & \forall v \in V,\\
& & & & x(\delta(S)) \geq 2, & \forall S\subset V, S \neq \emptyset,\\
& & & & 0 \leq x_{e} \leq 1, & \forall e \in E,  \elps where $\delta(S)$ is the set of all edges with exactly one endpoint in $S$ and we use the shorthand that $x(F) = \sum_{e \in F} x_e$.
This LP relaxation is sometimes called the {\em Subtour LP}.  It is not hard to show that given a feasible solution $x$ to the Subtour LP, $\frac{n-1}{n}x$ is feasible for the spanning tree polytope $\{x \in \Re^{|E|}: x(E) = n-1, x(E(S)) \leq |S|-1 ~~\forall S \subseteq V, |S| \geq 2\}$, where $E(S)$ is the set of all edges with both endpoints in $S$.  Similarly, any feasible solution $x$ to a related relaxation of the $s$-$t$ path TSP is also feasible for the spanning tree polytope.  Oveis Gharan, Saberi, and Singh \cite{OveisGharanSS11} propose an algorithm which has since been called (by \cite{AnKS12}) {\em Best-of-Many Christofides}: given the LP solution $x^*$, we can compute in polynomial time a decomposition of $x^*$ into a convex combination of spanning trees, and we run Christofides' algorithm (or the Hoogeveen variant for $s$-$t$ path TSP) for each of these trees and output the lowest cost solution found.  More precisely, if $\chi_F \in \{0,1\}^{|E|}$ is the characteristic vector of a set of edges $F$, then given Subtour LP solution $x^*$, we find spanning trees $F_1,\ldots,F_k$ such that $\frac{n-1}{n} x^* = \sum_{i=1}^k \lambda_i \chi_{F_i}$ for $\lambda_i \geq 0$ and $\sum_{i=1}^k \lambda_i = 1$.  Then for each tree $F_i$ we find a matching $M_i$ of the odd-degree vertices, and we compute a tour by shortcutting $F_i \cup M_i$.  We return the cheapest tour found.

An alternative perspective is to consider randomly sampling a spanning tree from a distribution on spanning trees given by the convex combination, then run Christofides' algorithm on the resulting tree found; that is, we sample tree $F_i$ with probability $\lambda_i$.  This perspective potentially allows us to avoid computing the convex combination explicitly.  However, the distribution of trees then depends on the (implicit) convex combination.  Asadpour et al.\ \cite{AsadpourGMOS10} and Oveis Gharan, Saberi, and Singh \cite{OveisGharanSS11} use a {\em maximum entropy} distribution.    For the Asadpour et al.\  ATSP result, the main property used of the maximum entropy distribution is that in some cuts of edges, the appearance of arcs is  negatively correlated.  Chekuri, Vondr\'ak, and Zenklusen \cite{ChekuriVZ10} show how to draw a sample with the appropriate negative correlation properties given an explicit convex combination of trees; their distribution over trees is not the same as the maximum entropy distribution.

An exciting possible direction for an improved approximation algorithm for the TSP is to show that some variation of the Best-of-Many Christofides' algorithm gives a performance guarantee strictly better than 3/2, either by starting with a convex combination of spanning trees, or using some of the stronger properties obtained by sampling a tree from a negatively correlated distribution, or the maximum entropy distribution.
In this paper, we experimentally evaluate these different versions of the Best-of-Many Christofides' algorithm in order to see if there is empirical evidence that these algorithmic variants are any better than the standard Christofides' algorithm, and whether any of the variants is more promising than the others.

In particular, we start by implementing Christofides' algorithm.  Since most of our instances are geometric, we compute a Delaunay triangulation using the package Triangle \cite{Shewchuk96}; it is known that the edges of an MST for a 2D Euclidean instance are a subset of the edges of the Delaunay triangulation.  We use Prim's algorithm to compute the MST from these edges.  For non-geometric instances, we use Prim's algorithm to compute the MST.  We then use the Blossom V code of Kolmogorov \cite{Kolmogorov09} to find a minimum-cost perfect matching on the odd degree vertices of the tree. We compute a tour by shortcutting the resulting Eulerian graph; we perform a simple optimization on the shortcutting.  We then use the Concorde TSP solver \cite{Concorde} to compute a solution $x$ to the subtour LP.  We implement two different ways of finding an explicit convex combination of trees equal to $\frac{n-1}{n}x$; in the first, we use a column generation technique suggested by An in his Ph.D.\ thesis \cite{An12} in conjunction with the linear programming solver Gurobi \cite{Gurobi}.  In the second, we compute a packing of spanning trees via iteratively ``splitting off'' edges of the LP solution from vertices, then maintaining a convex combination of trees as we ``lift back'' the split-off edges.  For this algorithm, we use a subroutine of Nagamochi and Ibaraki \cite{NagamochiI97} to obtain a complete splitting-off of a vertex.  We also implement two methods for obtaining a randomly sampled tree from the support of the LP solution.  We first implement the SwapRound procedure of Chekuri et al.\ \cite{ChekuriVZ10}; given an explicit convex combination of trees generated by the first two methods, we can sample a spanning tree such that the edges of the tree appearing in any given set are negatively correlated (we define the negative correlation more precisely in Section \ref{sec:swapround}).  We also implement the method for computing a maximum entropy distribution over spanning trees given the LP solution $x^*$, and then drawing a sample from this distribution, as given in the ATSP paper of Asadpour et al.\ \cite{AsadpourGMOS10} and the Ph.D.\ thesis of Oveis Gharan \cite{OveisGharan13}.  Our implementation choices for the maximum entropy routine were influenced by a code shared with us by Oveis Gharan \cite{OveisGharan14}.

To test our results, we ran these algorithms on TSPLIB instances of Reinelt  \cite{Reinelt91} (both Euclidean and non-Euclidean instances) and Euclidean VLSI instances from Rohe \cite{Rohe}.  We also considered graph TSP instances to see if the performance of the algorithms was better for such instances than for weighted instances. For our graph TSP instances, we used undirected graphs from the Koblenz Network Collection of Kunegis \cite{Kunegis13}.

It is known that the standard Christofides' algorithm typically returns solutions of cost of about 9-10\% away from the cost of an optimal solution on average (see, for instance, Johnson and McGeoch \cite{JohnsonM02}); this is better than its worst-case guarantee of at most 50\% away from the cost of an optimal solution, but not as good as other heuristics (such as the Lin-Kernighan heuristic \cite{LinK73}) that do not have performance guarantees.  We confirm these results for the standard Christofides' algorithm on geometric instances, but Christofides' algorithm appears to do worse on graph TSP instances; we had solutions of about 12\% away from optimal.  All of the Best-of-Many Christofides' algorithms performed substantially better than the standard Christofides' algorithm, with solutions of cost about 3-7\% away from optimal for the Euclidean instances, 2-3\%  away from optimal for the non-Euclidean instances, and under 1\% away from optimal for the graph TSP instances. These results may indicate that the graph TSP instances are easier for LP-based algorithms than geometric instances.  The algorithm that used the maximum entropy distribution on average outperformed the other Best-of-Many Christofides' algorithms; however, the algorithm that found a convex combination of spanning trees via splitting off, then used the SwapRound routine, was nearly as good as maximum entropy sampling, and was better in some cases.

A very recent example of Schalekamp and van Zuylen \cite{SchalekampvZ15} suggests that running the Best-of-Many Christofides' algorithm on  a fixed convex combination of trees cannot be better than a $\frac{3}{2}$-approximation algorithm; this may help explain why the random sampling methods perform better in practice.  We discuss this matter further in our concluding section.

Our paper is structured as follows.  In Section \ref{sec:algs}, we give a more detailed description of the algorithms that we implemented.  In Section \ref{sec:experiments}, we describe the TSP datasets we used and our machine environment, and in Section \ref{sec:results} we give the results of our experiments, as well as some analysis.  We conclude in Section \ref{sec:conc}.

\section{Algorithms}
\label{sec:algs}

In this section, we give descriptions of the various algorithms that we implemented.  Sections \ref{sec:colgen} and \ref{sec:splitting} describe the two algorithms that generate explicit convex combinations of spanning trees given the Subtour LP solution $x^*$; as described above, we compute the Subtour LP solution by using the Concorde TSP solver \cite{Concorde}.  Section \ref{sec:swapround} describes the SwapRound algorithm of Chekuri, Vondr\'ak, and Zenklusen \cite{ChekuriVZ10} that generates a randomly sampled tree given an explicit convex combination of spanning trees.  Section \ref{sec:maxentropy} describes the maximum entropy distribution on spanning trees, the algorithm used to compute it, and the algorithm used to draw a sample from this distribution.

As described in the introduction, we also have an implementation of the basic Christofides' algorithm, in which we start with the minimum-cost spanning tree given by Prim's algorithm; for the two-dimensional geometric instances, we first calculate the Delaunay triangulation using Triangle \cite{Shewchuk96} and run Prim's algorithm on these edges.  For the graph TSP instances, because the graphs are connected, the trees will always have a cost of $n-1$.  We added a very small random weight to each edge for these instances so that the algorithm chooses a random spanning tree of the graph.  We then find a minimum-cost perfect matching by using the Blossom V code of Kolmogorov \cite{Kolmogorov09}.  In our shortcutting, we build up a tour; if we encounter a vertex $v$ in the Eulerian traversal that is already part of the tour, we choose whether inserting $v$ at this point in the tour and removing it from the previous point would be cheaper than skipping over $v$ at this point in the tour and going on to the next vertex in the traversal.

\subsection{Column Generation}
\label{sec:colgen}

Our first algorithm for decomposing the Subtour LP solution $\frac{n-1}{n}x^*$ into a convex combination of spanning trees follows an algorithm described by An in his Ph.D.\ thesis \cite{An12}; we use column generation to generate the trees in the convex combination.  In particular, we would like to solve the following linear program, in which we have a variable $y_T$ for each possible spanning tree $T$, where we assume that $x^*_e$ is the given solution to the Subtour LP and the graph $G=(V,E)$ is on the edges $E= \{e: x^*_e > 0\}$:
\lps & &
& \mbox{Min} & \sum_{e \in E} s_{e} \\
& \mbox{subject to:} \\
& & & & \sum_{T: e \in T} y_T + s_e = \frac{n-1}{n}x^*_e, & \forall e \in E,\\
& & & & y_T \geq 0, & \forall T.\\
& & & & s_{e} \geq 0, & \forall e \in E.  \elps
Since $\frac{n-1}{n} x^*$ can be expressed as a convex combination of spanning trees, the optimal solution to the LP is zero, and for an optimal solution $(y^*,s^*)$, $y^*$ gives a convex combination of spanning trees.
The dual of this LP is:
\lps & &
& \mbox{Max} & \frac{n-1}{n} \sum_{e \in E} x^*_e z_e \\
& \mbox{subject to:} \\
& & & & \sum_{e \in T} z_e \leq 0, & \forall T,\\
& & & &  z_e \leq 1, & \forall e \in E. \elps
Because there are too many variables $y_T$ to include initially, we use a column generation approach.  We initially include some variables $y_T$ representing some spanning trees in the support of the $x^*_e$, solve the LP, and find the dual variables $z_e$.  We then compute a maximum-cost spanning tree $\hat T$ using the value $z_e$ as the cost of the edge $e$ for each edge $e \in E$.  If this cost $\sum_{e \in \hat T} z_e \leq 0$, then the current solution is optimal; otherwise, we add a new variable $y_{\hat T}$ corresponding to the tree $\hat T$.  We use Gurobi \cite{Gurobi} to solve the LPs and Prim's algorithm to compute the maximum-cost spanning trees.

\begin{figure}
\begin{center}
\begin{subfigure}{.49\textwidth}
  \centering
  \includegraphics[width=.8\linewidth]{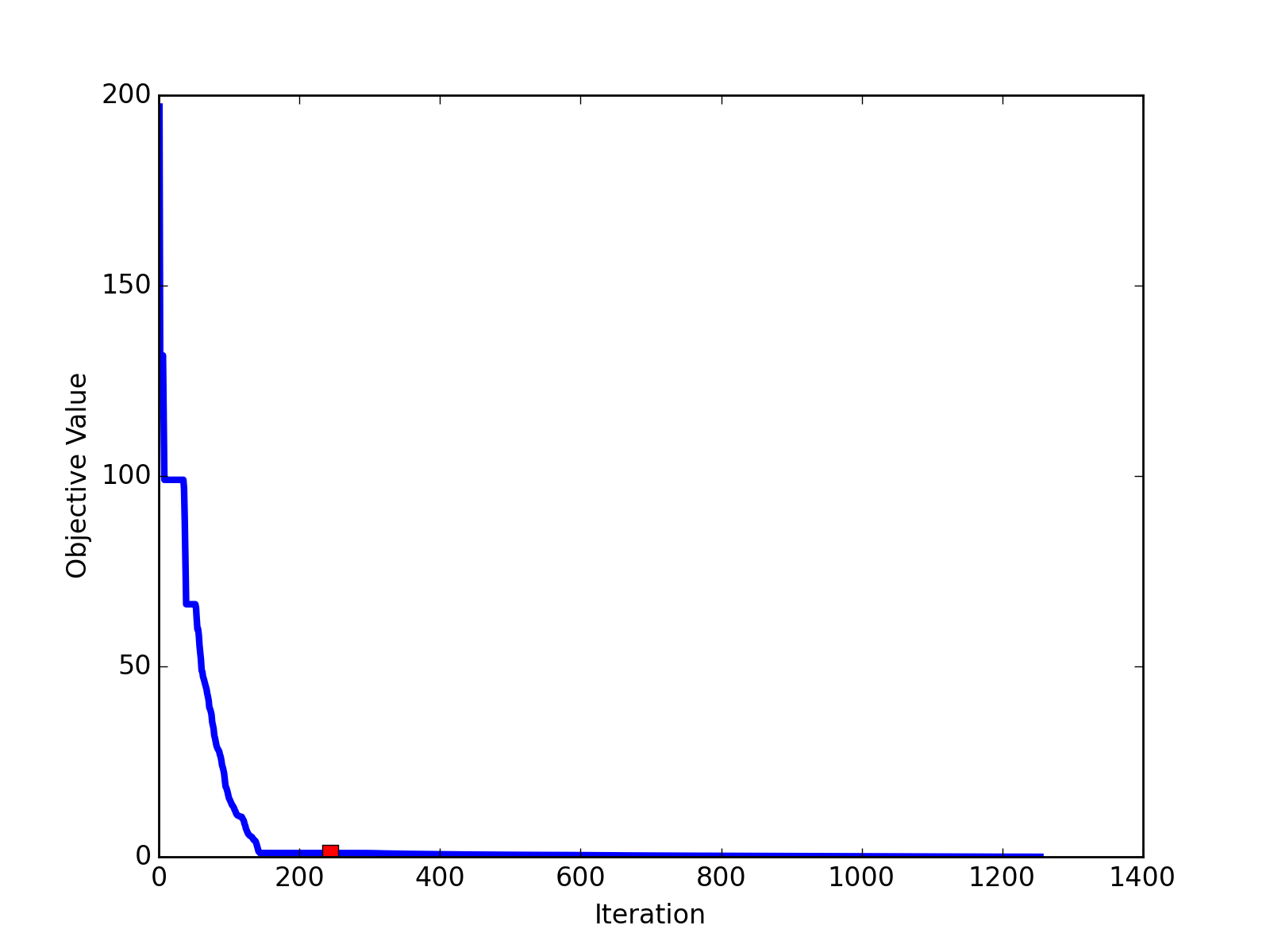}
  \caption{TSPLIB D198.}
\end{subfigure}
\begin{subfigure}{.49\textwidth}
  \centering
  \includegraphics[width=.8\linewidth]{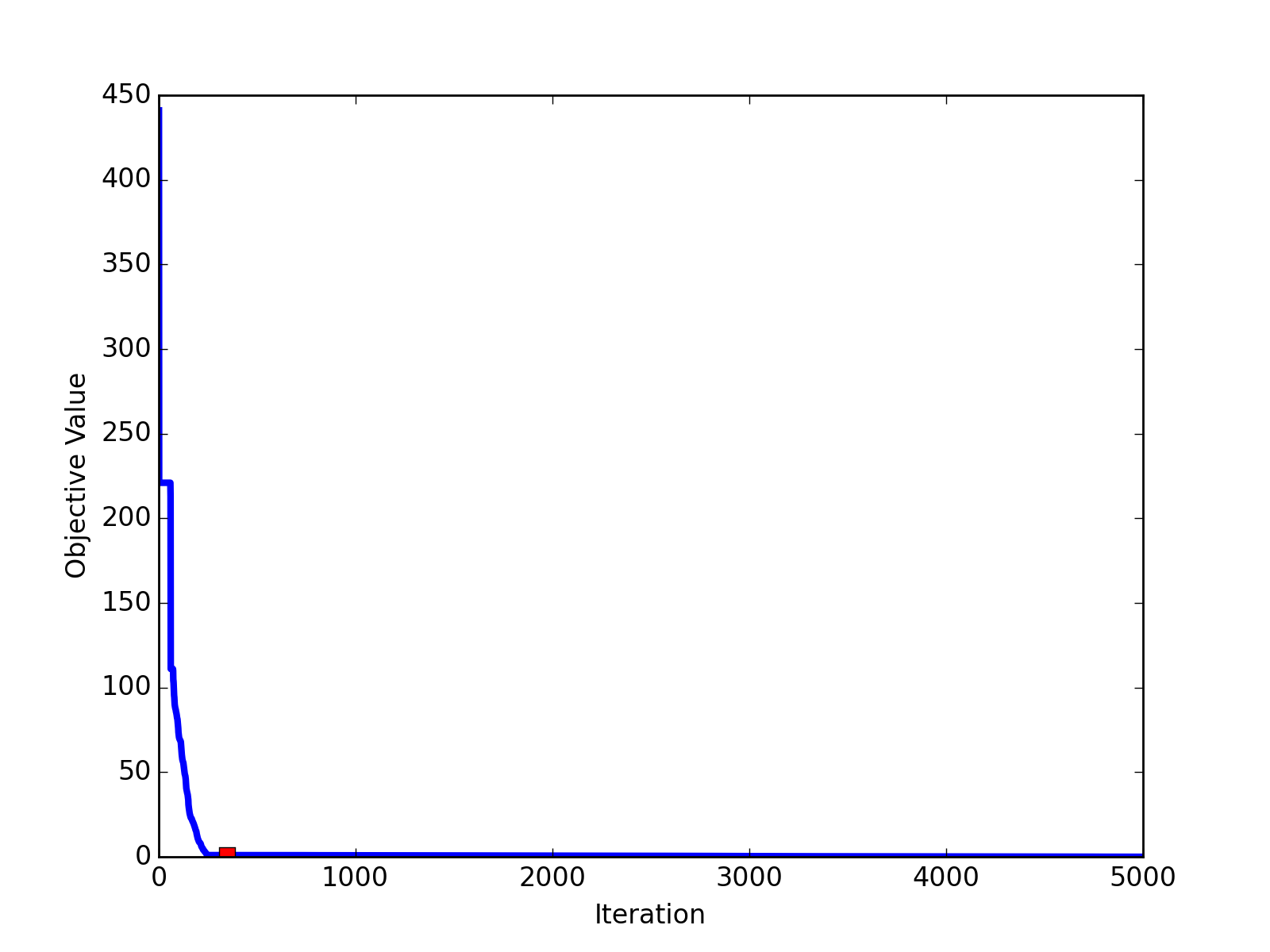}
  \caption{TSPLIB PCB442.}
\end{subfigure}
\end{center}
\caption{Column generation objective function value versus iteration count for some sample data sets, run both with and without our cutoff procedure.  The small rectangle indicates the iteration in which our cutoff procedure terminates the column generation algorithm early.}
\label{fig:colgen}
\end{figure}

As is typical of column generation algorithms, solving the LP to optimality can take a long time.  On instances of 500 cities, it could take up to 10 hours of computation time.  Thus in order to make comparisons with other methods, we terminated early.  In particular, we store the current objective function value and wait until either the value drops by .1 or 100 iterations have occurred.  If the objective has not dropped by .1 in the 100 iterations, we terminate; otherwise, once it has dropped by .1 we restart the iteration count.  Our cutoff behavior allows us to avoid the long set of iterations in which the method is near optimal but only makes incremental progress; see Figure \ref{fig:colgen} for an illustration of how the cutoff helps for two instances.

\subsection{Splitting Off and Tree Packing}
\label{sec:splitting}

Let $x^*$ be a basic feasible solution to the Subtour LP.  It is known that $x^*$ is rational, so that there exists a $K$ such that $Kx^*$ is integer.  We let $\hat G$ be a multigraph with $Kx^*_e$ copies of edge $e$.  Then by the constraints of the Subtour LP, each vertex has $2K$ edges incident on it, and for each $i, j \in V$, there are at least $2K$ edge-disjoint paths between $i$ and $j$.  Lov\'asz \cite{Lovasz76} showed that given any edge $(x,z)$ incident on $z$ it is possible to find another edge $(y,z)$ incident on $z$ such that if we remove $(x,z)$ and $(y,z)$ from $\hat G$ and add edge $(x,y)$, then for all pairs of vertices $i,j \in V$, $i,j \neq z$, there are still $2K$ edge-disjoint paths between $i$ and $j$.  Removing $(x,z)$ and $(y,z)$ and adding $(x,y)$ so as to preserve connectivity in this way is called the {\em splitting off} operation.  A {\em complete splitting off} at $z$ removes all $2K$ edges incident at $z$ from $\hat G$ and adds $K$ edges not incident on $z$ to $\hat G$ so that for all pairs of vertices $i,j \in V$, $i,j \neq z$, there are still $2K$ edge-disjoint paths between $i$ and $j$.  Nagamochi and Ibaraki \cite{NagamochiI97} give an $O(nm\log n + n^2 \log^2 n)$ time algorithm for obtaining a complete splitting off at a node.

We implemented the Nagamochi-Ibaraki algorithm, and use it to obtain a set of $K$ trees in the original multigraph $\hat G$ as follows (following Frank \cite[Chapter 10]{Frank11}).  We first iterate through all but two nodes of the graph, obtain a complete splitting off at each node, and keep track of the splitting off in each case (that is, we remember that edge $(x,y)$ results from splitting off $(x,z)$ and $(y,z)$).  We then inductively construct a set of $K$ trees, starting with the two nodes, and adding back a node at a time in reverse of the order in which we did the splitting off operations.  The base case is simple: for the two nodes, there will be $2K$ edges remaining that join them.  We choose $K$ edges arbitrarily, so that we now have $K$ spanning trees on the two nodes.  The remaining edges we call {\em leftover edges}.  In the inductive step, we have $K$ trees spanning a set of nodes and $K$ leftover edges; we wish to extend the trees to also span a new node $z$ by undoing the splitting off operations from node $z$.  There are $K$ edges that were split off from $z$; these $K$ edges appear in the current set of trees or in the set of leftover edges.  We maintain a set of {\em leaf edges} $F$ in the step.  We start by considering any split off edge $(x,y)$ that is a leftover edge; we replace $(x,y)$ with $(x,z)$ in the set of leftover edges, and add $(y,z)$ to the set of leaf edges.  For any tree that contains at least one split off edge $(x,y)$, we remove $(x,y)$ from the tree and add $(x,z)$ and $(z,y)$ to the tree.  If the same tree contains other split off edges $(x',y')$, then for each such edge we calculate the shortest path distance in the tree from $z$ to all other nodes in the tree.  We remove $(x',y')$ and add either $(x',z)$ or $(y',z)$, depending on whether $x'$ or $y'$ is farther away from $z$ (note that because it is a tree, they cannot be at the same distance); adding the edge with the farther endpoint from $z$ maintains the connectivity of the tree.  The other edge that is not added to the tree ($(x',z)$ or $(y',z)$) we add to the set of leaf edges.  Finally, for each tree that contains no split off edges, we remove an arbitrary edge from the set of leaf edges and add it to the tree.  At the end of the process for node $z$, we have $K$ trees that also span $z$, have no remaining leaf edges, and have exactly $K$ leftover edges.  Once we have iterated through all the nodes, we will have a set of $K$ trees and $K$ leftover edges such that if $\chi_{F_i}$ is the characteristic vector of the $i$th tree and $\chi_L$ is the characteristic vector of the multiset of leftover edges, then $\frac{1}{K} \sum_{i=1}^K \chi_{F_i} + \frac{1}{K} \chi_L  = x^*$.  Thus $\frac{1}{K} \sum_{i=1}^K \chi_{F_i} \leq x^*$.

\subsection{SwapRound and Negatively Correlated Distributions}
\label{sec:swapround}

The algorithms of the previous two sections give a convex combination of spanning trees such that the convex combination of the characteristic vectors of the trees is dominated by the Subtour LP solution $x^*$.  Let $z^* = \sum_{i=1}^K \lambda_i \chi_{F_i} \leq x^*$ give the convex combination of the characteristic vectors $\chi_{F_i}$ of $K$ trees $F_i$.  One can think about the convex combination as being a distribution on spanning trees: we sample tree $F_i$ with probability $\lambda_i$.  A nice feature of this sampling scheme is that the expected cost of the sampled tree is $\sum_{e \in E} c_e z^*_e \leq \sum_{e \in E} c_e x^*_e$, at most the value of the Subtour LP.  This follows since the probability that a given edge $e \in E$ is in the sampled tree $F$ is $$\Pr[e \in F] = \sum_{i: e \in F_i} \lambda_i = \sum_{i: e \in F_i} \lambda_i \chi_{F_i}(e) = z^*_e \leq x^*_e.$$  Let $X_e$ be a random variable which is 1 if edge $e$ is in the sampled tree and 0 otherwise; then we have shown that $E[X_e] = z^*_e$.

Asadpour et al.\ \cite{AsadpourGMOS10} show that for proving results about the asymmetric TSP, it is useful to think about drawing a sample such that the edges of the spanning tree appearing in a fixed set are {\em negatively correlated}.  We will say that a probability distribution is negatively correlated if $E[X_e] = z^*_e$ and for any set of edges $A \subseteq E$, $E\left[\Pi_{e \in A} X_e\right] \leq \prod_{e \in A} z^*_e$, and $E\left[\Pi_{e \in A} (1-X_e) \right] \leq \prod_{e \in A} (1-z^*_e)$.  Negative correlation allows the proof of concentration bounds that get used in the result of Asadpour et al.

Chekuri, Vondr\'ak, and Zenklusen \cite{ChekuriVZ10} give a sampling scheme they call SwapRound such that given any convex combination of spanning trees as input, SwapRound gives a sample from a negatively correlated distribution as output (their result applies more generally to matroids).  Let $F_1,\ldots,F_k$ be the trees from the convex combination.  The algorithm maintains a spanning tree $F$, which is initially $F_1$.  Then it loops through the other trees $F_2,\ldots,F_k$, and calls a subroutine, MergeBasis, with the two trees $F$ and $F_i$, and with probability weights $\sum_{j=1}^{i-1} \lambda_j$ and $\lambda_i$, and updates $F$ to be the result of MergeBasis.  The routine MergeBasis, given two trees $F$ and $F'$ and weights $\lambda$ and $\lambda'$, repeatedly interchanges edges between the trees $F$ and $F'$ until the two are the same.  While $F \neq F'$, the routine finds edges $e \in F-F'$ and $e' \in F'-F$ such that $F-e+e'$ is a spanning tree and $F'-e'+e$ is a spanning tree (such edges are known to exist if $F\neq F'$).  Then with probability $\lambda/(\lambda+\lambda')$, the routine updates $F'$ to $F'-e'+e$, and otherwise the routine sets $F$ to $F-e+e'$.  When $F=F'$, the routine returns $F$.

We implemented the SwapRound and MergeBasis routines in order to see if sampling a spanning tree from a negatively correlated distribution would lead to better overall results. Because both the column generation and splitting off methods give a convex combination of spanning trees, we tried these two methods both with and without the SwapRound routine on the output.  Because the sampling can be performed in parallel, we had four threads running to draw the samples.  We drew 1000 samples per instance, and output the best tour found.

\subsection{The Maximum Entropy Distribution}
\label{sec:maxentropy}

Asadpour et al.\  \cite{AsadpourGMOS10} consider sampling spanning trees from the maximum entropy distribution over spanning trees.  Given the subtour LP solution $x^*$, we set $z^* = \frac{n-1}{n}x^*$.  If ${\cal T}$ is the set of all spanning trees of the graph $G$, then the maximum entropy distribution is an optimal solution to the following:
\lps & &
& \mbox{Inf} & \sum_{T \in {\cal T}} p(T) \log p(T) \\
& \mbox{subject to:} \\
& & & & \sum_{T: e \in T} p(T) = z^*_e, & \forall e \in E,\\
& & & & \sum_{T \in {\cal T}} p(T) = 1\\
& & & & p(T) \geq 0, & \forall T.  \elps
Asadpour et al.\ show that the constraint $\sum_{T \in {\cal T}} p(T) = 1$ is redundant.  Given that $z^*$ is in the relative interior of the spanning tree polytope, they argue that there must exist $\gamma^*_e$ for all $e \in E$ such that sampling tree $T$ with probability proportional to $p(T) = e^{\gamma^*(T)}$ (with $\gamma^*(T) \equiv \sum_{e \in T} \gamma^*_e$) results in $\Pr[e \in T] = z^*_e$ and gives the maximum entropy distribution.

Asadpour et al.\ then give an algorithm for computing values $\tilde{\gamma_e}$ that approximately satisfy the conditions.  In particular, the value $\tilde{\gamma_e}$ are such that if we set
$$\tilde{p}(T) \equiv \frac{1}{P} \mathrm{exp}\left(\sum_{e \in T}\tilde{\gamma_e}\right), \mbox{ where } P\equiv \sum_{T \in {\cal T}} \mathrm{exp} \left(\sum_{e \in T} \tilde{\gamma_e}\right),$$
then
$$\tilde{z_e} \equiv \sum_{T \in {\cal T} : T \ni e} \tilde{p}(T) \leq (1 + \epsilon)z^*_e.$$  To compute the $\tilde{\gamma_e}$, we use a combination of the algorithm suggested in Asadpour et al.\ and one given by code written by Oveis Gharan \cite{OveisGharan14}.  We use the Eigen C++ template library \cite{Eigen} to perform matrix operations needed by the code (e.g.\ sparse matrix storage, matrix-vector multiplication, and matrix inversion).  As part of the computation, we need to compute the inverse of a weighted Laplacian matrix.  The Laplacian matrix $L=(\ell_{ij})$ of a undirected graph with weights $w_e$ for all $e \in E$ has $\ell_{ij} = \ell_{ji} = -w_e$ for $e=(i,j)$, and $\ell_{ii} = \sum_{j: (i,j) \in E} \ell_{ij}.$  The algorithm we use is as follows, where we set $\epsilon = 0.01$:
\begin{enumerate}
	\item
	Set all $\tilde{\gamma}_e$ values to 0, and let $z_e = z^*_e$.
	\item
	Define $q_e(\tilde{\gamma})$ to be the probability that an edge $e = (i,j)$ will be included in a spanning tree $T \in {\cal T}$ that is sampled with probability proportional to $\exp\left(\sum_{e \in T} \tilde\gamma_e \right)$.   We iteratively improve $\bf \tilde{\gamma}$ until there is no $\tilde{\gamma_e}$ value such that  $q_e(\tilde \gamma) > (1 + \epsilon)z_e$, according the following procedure:
	\begin{itemize}
		\item
		Estimate each $q_e(\tilde \gamma)$ for $e=(i,j)$ using the current $\tilde{\gamma_e}$ values by first creating the weighted Laplacian matrix $L$ of the graph where each edge has weight $e^{\tilde{\gamma_e}}$ and inverting it. Then, calculate $q_e(\tilde \gamma)$ as $\vec{x}^{\mathsmaller{T}}L^{\mathsmaller{-1}}\vec{x}$, where $\vec{x}$ is the vector with the $i$th and $j$th entries as 1 and $-1$ respectively, and zeroes elsewhere.
		\item
		Loop over $\bf \tilde{\gamma}$, subtracting $\delta_e$ from each edge $e$ such that $q_e(\tilde \gamma) > (1 + \epsilon) z_e$, where
		$$\delta_e := \mathrm{log}  \left( \dfrac{ q_e(\tilde \gamma) \left(1 - \left(1+\dfrac{\epsilon}{2}\right)z_e\right)}{(1 - q_e(\tilde \gamma))\left(1+\dfrac{\epsilon}{2}\right)z_e} \right).$$
	\end{itemize}
\end{enumerate}

Once the $\tilde{\gamma_e}$ have been computed, we need to be able to sample from the corresponding distribution.  Asadpour et al.\ set $\lambda_e = e^{{\tilde \gamma_e}}$, and then sample a tree with probability proportional to $\prod_{e \in T} \lambda_e$.  Asadpour et al.\  give an algorithm for computing such a sample in polynomial time, which we implemented.  We also implemented the following algorithm for sampling $\lambda$-random trees used in the code of Oveis Gharan \cite{OveisGharan14}: we pick an arbitrary starting node $i$ as a location, and start with an empty edge set which will become the spanning tree. With probability proportional to the $\lambda_e$ for each edge $e=(i,j)$ incident on the current node $i$, pick an incident edge. If the node $j$ has not yet been visited, add edge $(i,j)$ to the tree. The new location becomes node $j$. Repeat until every node has been added to the tree (and hence the tree is spanning).
This algorithm is not guaranteed to find a sample in polynomial time; however, we found that it scaled to larger instances better than the algorithm given in Asadpour et al.\

\begin{figure}
\begin{subfigure}{.5\textwidth}
  \centering
  \includegraphics[width=.8\linewidth]{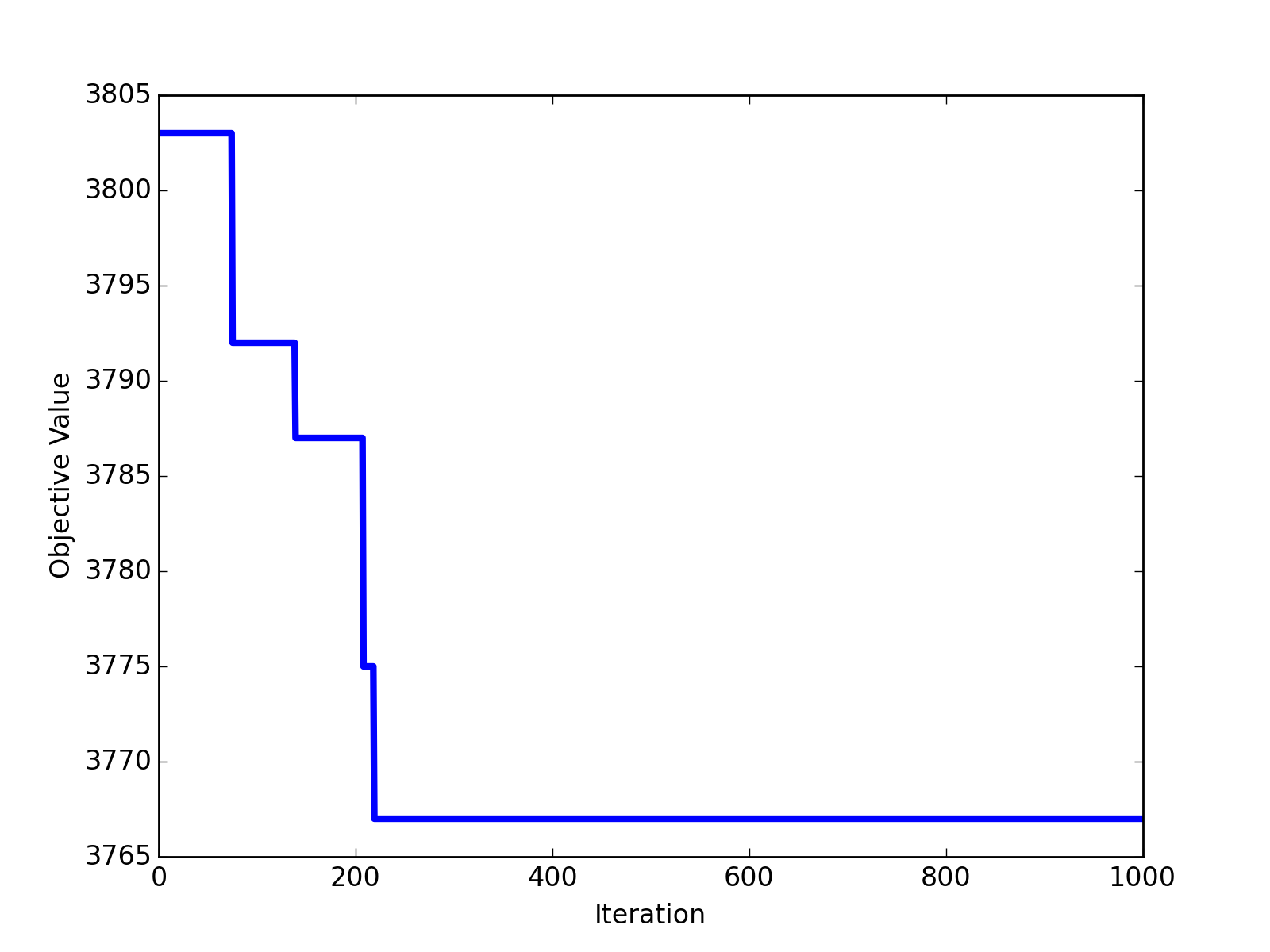}
  \caption{Maximum Entropy Sampling}
\end{subfigure}
\begin{subfigure}{.5\textwidth}
  \centering
  \includegraphics[width=.8\linewidth]{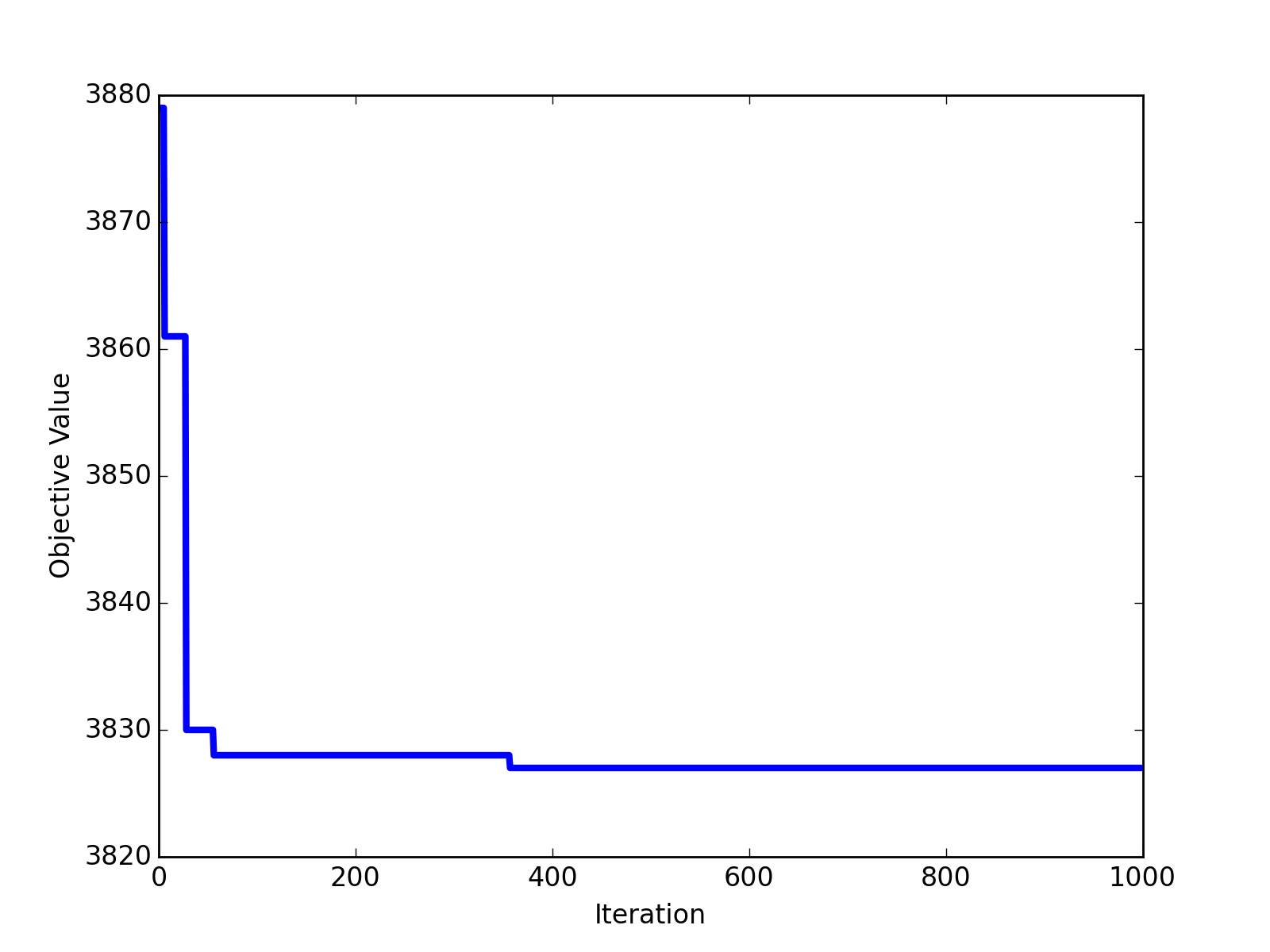}
  \caption{Column generation + SwapRound}
\end{subfigure}
\begin{center}
\begin{subfigure}{.5\textwidth}
  \centering
  \includegraphics[width=.8\linewidth]{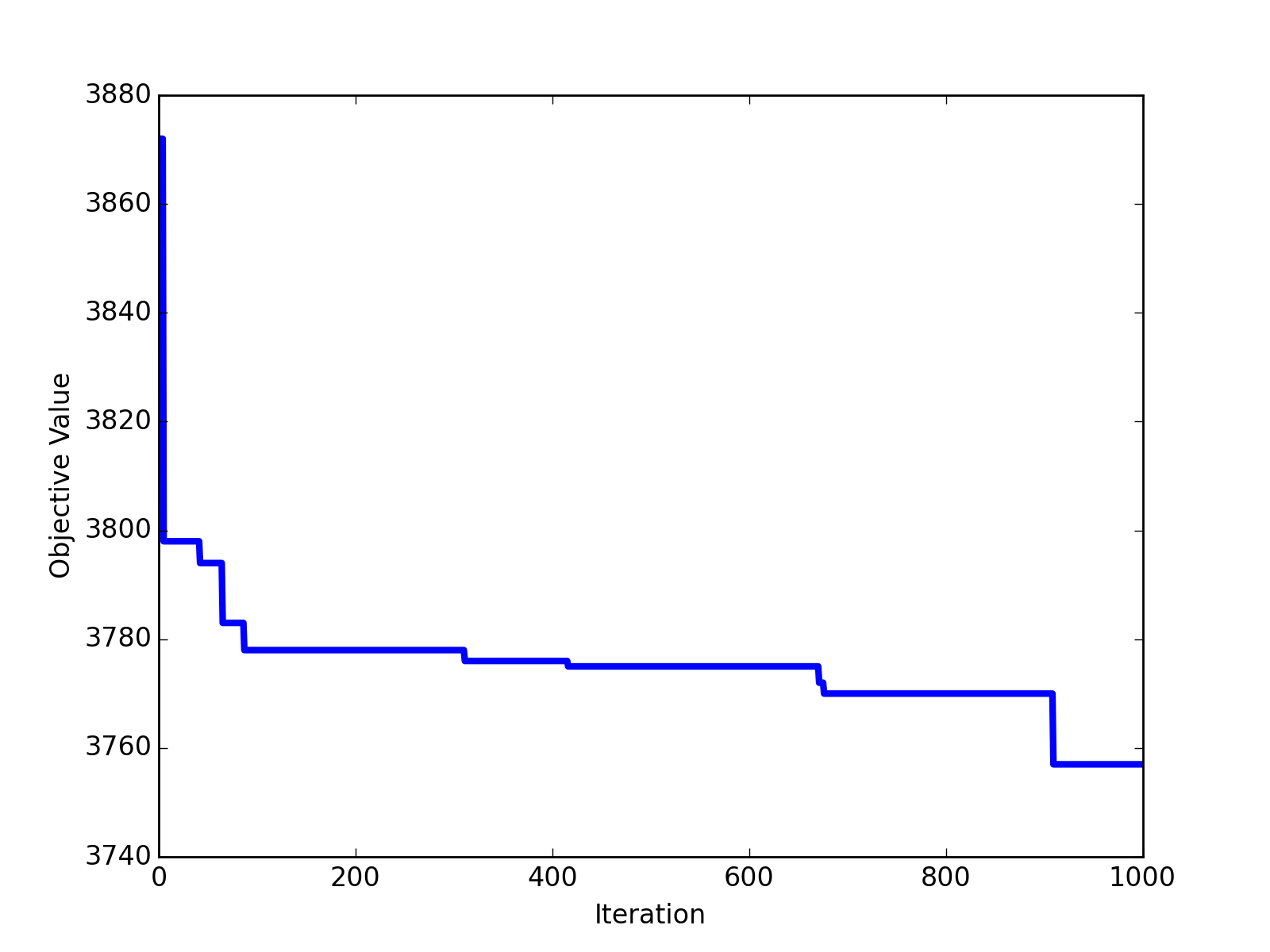}
  \caption{Splitting Off + SwapRound}
\end{subfigure}
\end{center}
\caption{Tradeoff between the number of samples drawn and the cost of the best tour found so far for the three different sampling methods on the VLSI instance XIT1083.}
\label{fig:sample}
\end{figure}

We  sampled trees as desired from the graph; as with the SwapRound method, 1000 samples were used for the results presented here.  For each sampled tree, we ran Christofides' algorithm, and output the lowest cost result found across all samples.  Once the $\tilde{\gamma}$ are computed, we can draw the samples in parallel using four threads.  The use of 1000 samples was not tuned, and we will consider another means for determining the number of samples to draw in the future.  In Figure \ref{fig:sample}, we can see that in some cases drawing that many samples is not particularly useful in improving the outcome (although in one case it is).

\section{Experiments}
\label{sec:experiments}

The algorithms above were implemented in C++.   We ran the algorithms on a machine with a 4.00Ghz Intel i7-875-K processor with 8GB DDR3 memory.

For test data, we used the TSPLIB instances of Reinelt \cite{Reinelt91}; we considered 59 two-dimensional Euclidean instances with up to 2103 vertices (averaging 524 vertices), and 5 non-Euclidean instances with 120 to 1032 vertices (gr120, si175, si535, pa561, and si1032), averaging 484 vertices. Similarly, we considered 39 two-dimensional Euclidean VLSI instances of Rohe \cite{Rohe} with up to 3694 vertices, averaging 1473 vertices.  We also considered instances of the graph TSP: we used 9 instances from the Koblenz Network Collection of Kunegis \cite{Kunegis13}.  Specifically, we considered undirected simple graphs; if the graph had multiple connected components, we used the largest connected component and discarded the rest of the graph.  The resulting instances ranged in size from 18 to 1615 vertices, averaging 363.  Because these instances did not have optimal solutions previously computed for them, we used Concorde \cite{Concorde} to compute the optimal.

\section{Results}
\label{sec:results}

A summary of our results can be found in Table \ref{tab:summary}.  It shows the percentage above optimal for the algorithms mentioned averaged across the four different data sets we used, the Euclidean TSPLIB instances of Reinelt \cite{Reinelt91}, the Euclidean VLSI instances of Rohe \cite{Rohe}, the non-Euclidean TSPLIB instances, and the graph TSP instances from the Koblenz Network Collection \cite{Kunegis13}, as described earlier. For the two methods that construct an explicit convex combination of spanning trees (column generation and splitting off), the best error is the error of the minimum-cost tour resulting from running Christofides' algorithm over all the trees in the decomposition, while the average error is the average error from running Christofides' algorithm over all the trees in the decomposition.  For the methods that sample trees from a distribution (the swap round variants and the maximum entropy distribution), the best error is the smallest error found over all tours resulting from running Christofides' algorithm on all the trees sampled from the distribution, and the average error is the average over all sampled trees.   We make several observations based on this summary.

\begin{table}
{\footnotesize
\begin{tabular}{| l | r | r r | r r | r r | r r | r r |}\hline
& \multicolumn{1}{|c|}{Std} & \multicolumn{2}{|c|}{ColGen} & \multicolumn{2}{|c|}{ColGen+SR} &\multicolumn{2}{|c|}{MaxEnt} & \multicolumn{2}{|c|}{Split} & \multicolumn{2}{|c|}{Split+SR} \\
& & \multicolumn{1}{|c}{Best} & \multicolumn{1}{c|}{Ave} & \multicolumn{1}{|c}{Best} & \multicolumn{1}{c|}{Ave} & \multicolumn{1}{|c}{Best} & \multicolumn{1}{c|}{Ave} & \multicolumn{1}{|c}{Best} & \multicolumn{1}{c|}{Ave} & \multicolumn{1}{|c}{Best} & \multicolumn{1}{c|}{Ave}  \\ \hline
TSPLIB (E) & 9.56\% &     4.03\% & 6.44\% & 3.45\% & 6.24\% &  3.19\% & 6.12\% & 5.23\% & 6.27\% & 3.60\% & 6.02\% \\
VLSI & 9.73\%   &  7.00\% & 8.51\% & 6.40\% & 8.33\% & 5.47\% & 7.61\% & 6.60\% & 7.64\% & 5.48\% & 7.52\% \\
TSPLIB (N)  & 5.40\% & 2.73\% & 4.41\% & 2.22\% & 4.08\% & 2.12\% & 3.99\% & 2.92\% & 3.77\% & 1.99\% & 3.82\%  \\
Graph & 12.43\% &  0.57\% & 1.37\% & 0.39\% & 1.29\% & 0.31\% & 1.23\% & 0.88\% & 1.77\% & 0.33\% & 1.20\% \\ \hline
\end{tabular}
}
\caption{Summary of results, giving the percentage in excess of optimal for the algorithms. `Std' is Christofides algorithm, `ColGen' is column generation, `MaxEnt' is maximum entropy sampling, `Split' is the splitting-off algorithm, and `SR' is the swap round algorithm.  The TSPLIB E instances are two-dimensional Euclidean, and the TSPLIB N instances are non-Euclidean. }
\label{tab:summary}
\end{table}

The first is that our results for Christofides' algorithm are very similar to those found by Johnson and McGeoch \cite{JohnsonM02}, at least for the Euclidean TSPLIB and VLSI instances, with roughly 9-10\% error; the error is less on the non-Euclidean TSPLIB instances.  Somewhat surprisingly, Christofides' algorithm seems to perform significantly worse on the graph TSP instances from the Koblenz Network Collection, with 12\% error.

One reason that the performance of Christofides' algorithm on the graph TSP instances is surprising is that for the other algorithms, the graph TSP instances seem to be significantly easier, with error under 1\%.  The VLSI instances appear to be the hardest overall for the algorithms collectively, but this may be because the average instance size is larger.

Another observation is that using SwapRound to sample trees does improve the overall performance of the output.

Of all the algorithms, drawing from the maximum entropy distribution gives the best overall results, but constructing the convex combination via splitting off and then applying SwapRound was quite close in most cases, and better in some.  Column generation is the worst of the variants, but we did not check whether the early termination of the column generation routine contributed to the weak performance of this variant.

Why are the Best-of-Many Christofides' algorithm variants significantly better than Christofides' algorithm?  The key is that they trade off significantly higher spanning tree cost against significantly lower matching costs, with the reduction in the matching costs outweighing the increase in the spanning tree cost; these results are summarized in Table \ref{tab:match}.  The average tree cost for all of the Best-of-Many Christofides' algorithm variants is at most the value of the subtour LP; the subtour LP is known to be very close to the cost of the optimal tour experimentally (about 98\%-99\% of optimal), and our experiments confirm this, while the minimum-cost spanning trees are 79\%-93\% of the cost of the optimal tour in our experiments.  However, the matching costs are dramatically reduced.  For Christofides' algorithm, the cost of the matching is 25\%-40\% of the cost of the optimal tour, while for the Best-of-Many Christofides' variants, it is 10\%-15\% in the case of the TSPLIB/VLSI instances, and 4-5\% for the graph TSP instances.

\begin{table}
{\small
\begin{tabular}{| l | r | r || r | r | r | r | r | r |}\hline
& \multicolumn{2}{|c||}{Tree} & \multicolumn{6}{|c|}{Matching} \\ \hline
& \multicolumn{1}{|c|}{Std} & \multicolumn{1}{|c||}{BOM} & \multicolumn{1}{|c|}{Std} & \multicolumn{1}{|c|}{ColGen} & \multicolumn{1}{|c|}{ColGen+SR} & \multicolumn{1}{|c|}{MaxEnt} & \multicolumn{1}{|c|}{Split} & \multicolumn{1}{|c|}{Split+SR} \\ \hline
TSPLIB  (E) & 87.47\% & 98.57\% & 31.25\% & 11.43\% & 11.03\% & 10.75\% & 10.65\% & 10.41\% \\
VLSI   & 89.85\% & 98.84\%  & 29.98\% & 14.30\% & 14.11\% & 12.76\% & 12.78\% & 12.70\% \\
TSPLIB (N) & 92.97\% & 99.36\% &  24.15\% & 9.67\% & 9.36\% & 8.75\% & 8.77\% & 8.56\%  \\
Graph  & 79.10\% &  98.23\% &  39.31\% & 5.20\% & 4.84\% & 4.66\% & 4.34\% & 4.49\% \\ \hline
\end{tabular}
}
\caption{Costs of trees and matchings for the various methods, all expressed relative to the cost of the optimal tour.}
\label{tab:match}
\end{table}

\begin{figure}
\begin{subfigure}{.5\textwidth}
  \centering
  \includegraphics[width=.8\linewidth]{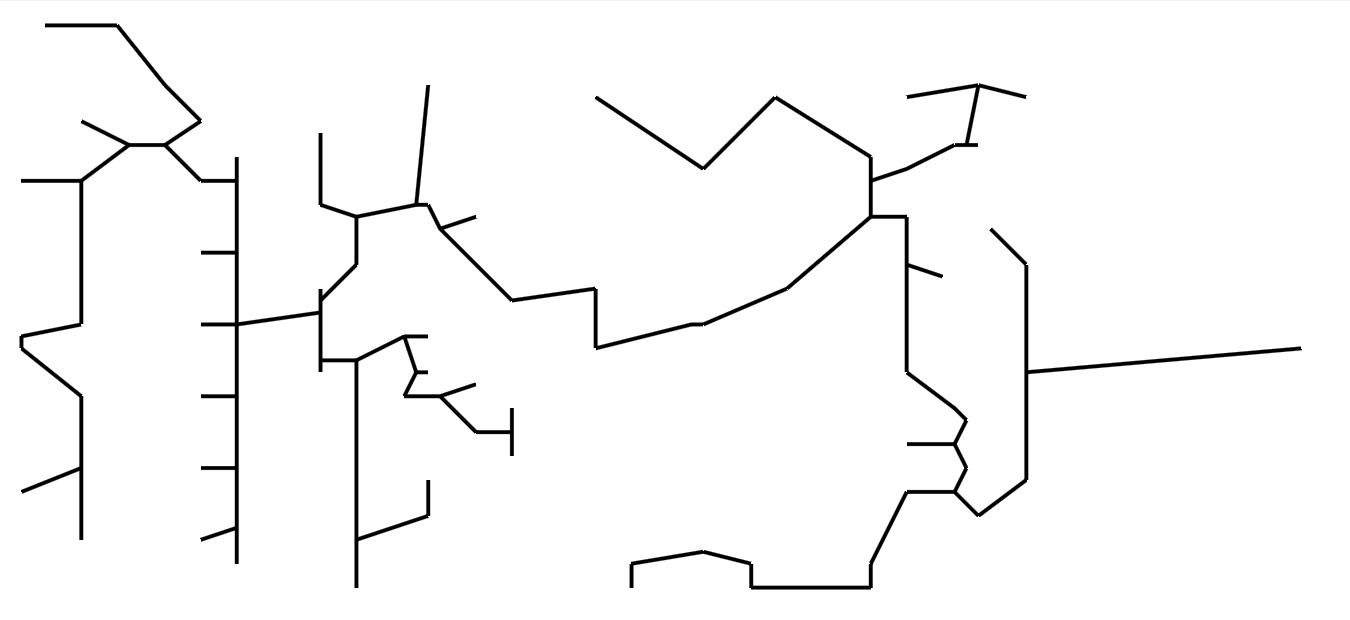}
  \caption{Standard Christofides MST}
\end{subfigure}%
\begin{subfigure}{.5\textwidth}
  \centering
  \includegraphics[width=.8\linewidth]{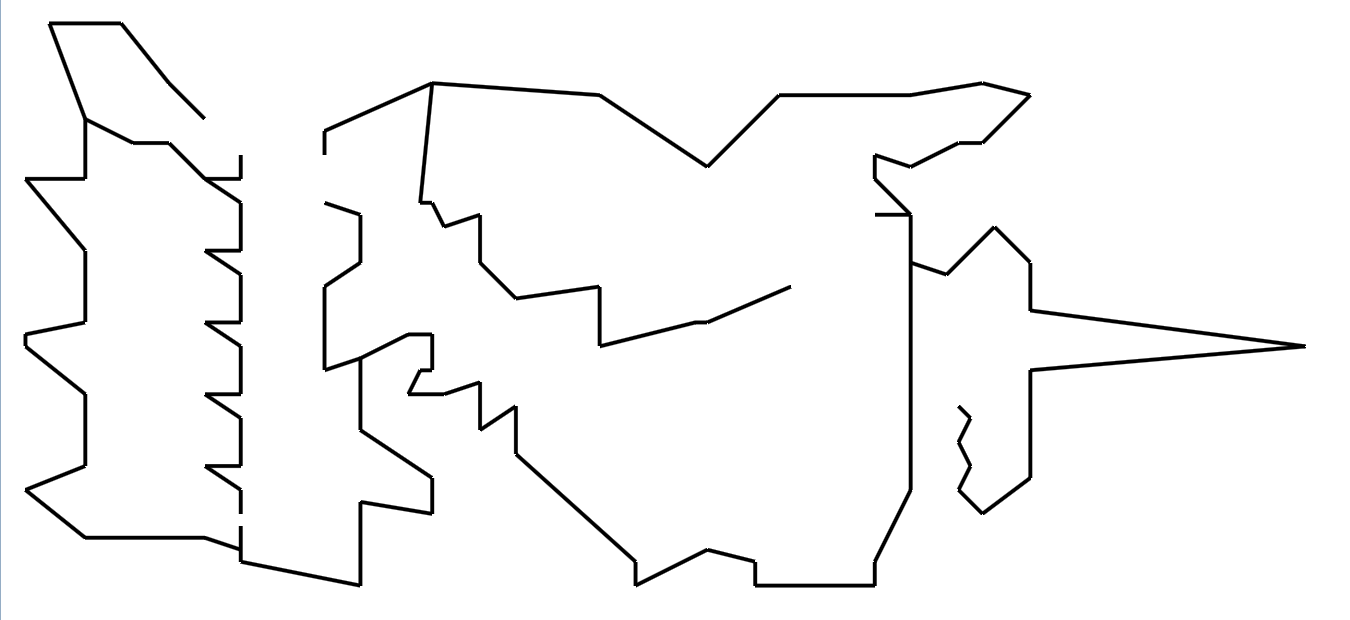}
  \caption{Column generation}
\end{subfigure}
\bigskip
\begin{subfigure}{.5\textwidth}
  \centering
  \includegraphics[width=.8\linewidth]{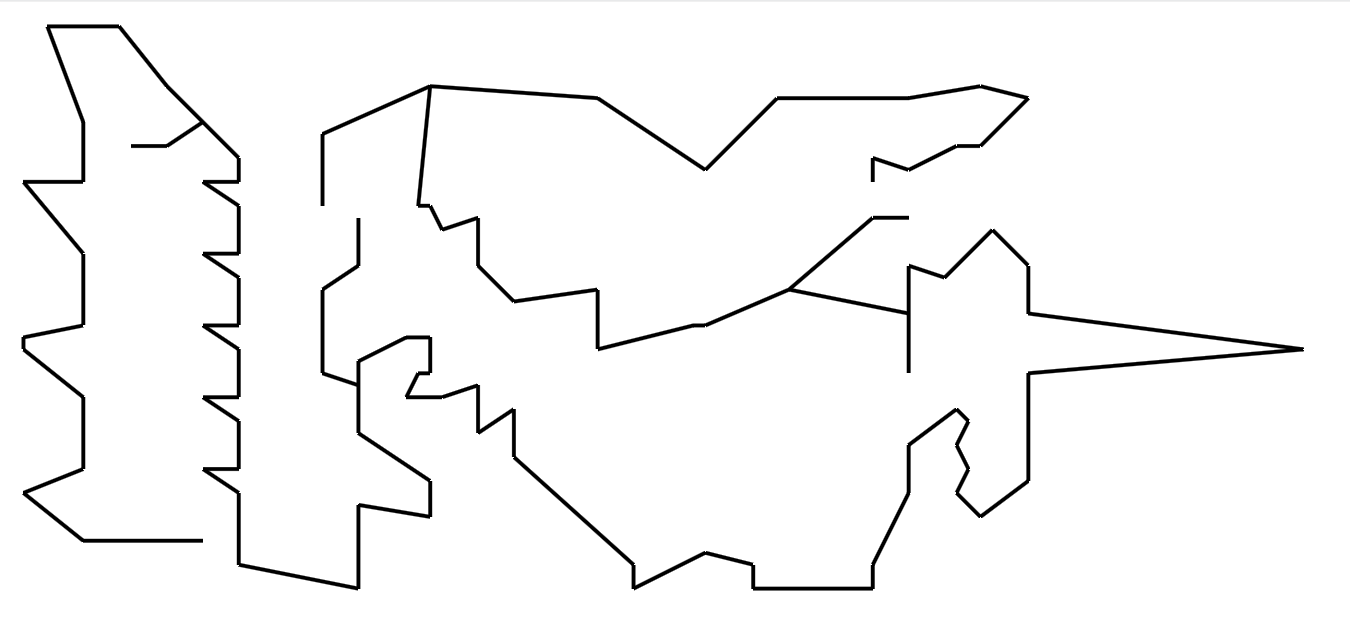}
  \caption{Column generation + SwapRound}
\end{subfigure}%
\begin{subfigure}{.5\textwidth}
  \centering
  \includegraphics[width=.8\linewidth]{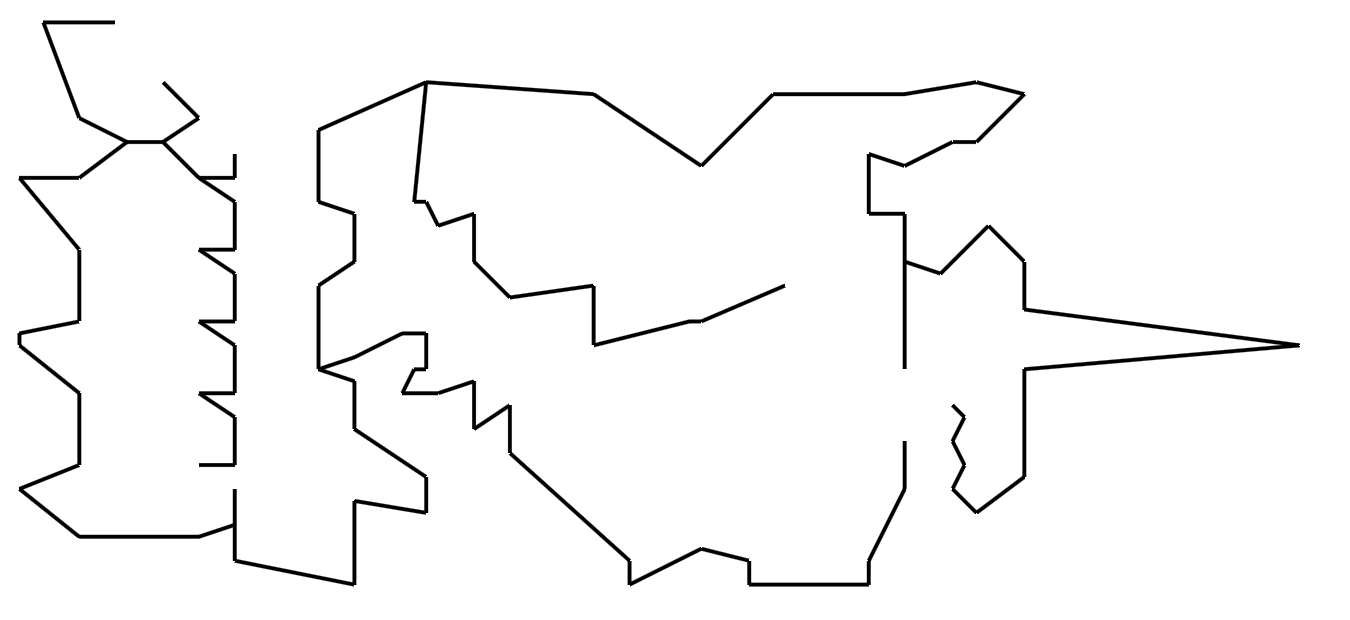}
  \caption{Maximum entropy}
\end{subfigure}
\bigskip
\begin{subfigure}{.5\textwidth}
  \centering
  \includegraphics[width=.8\linewidth]{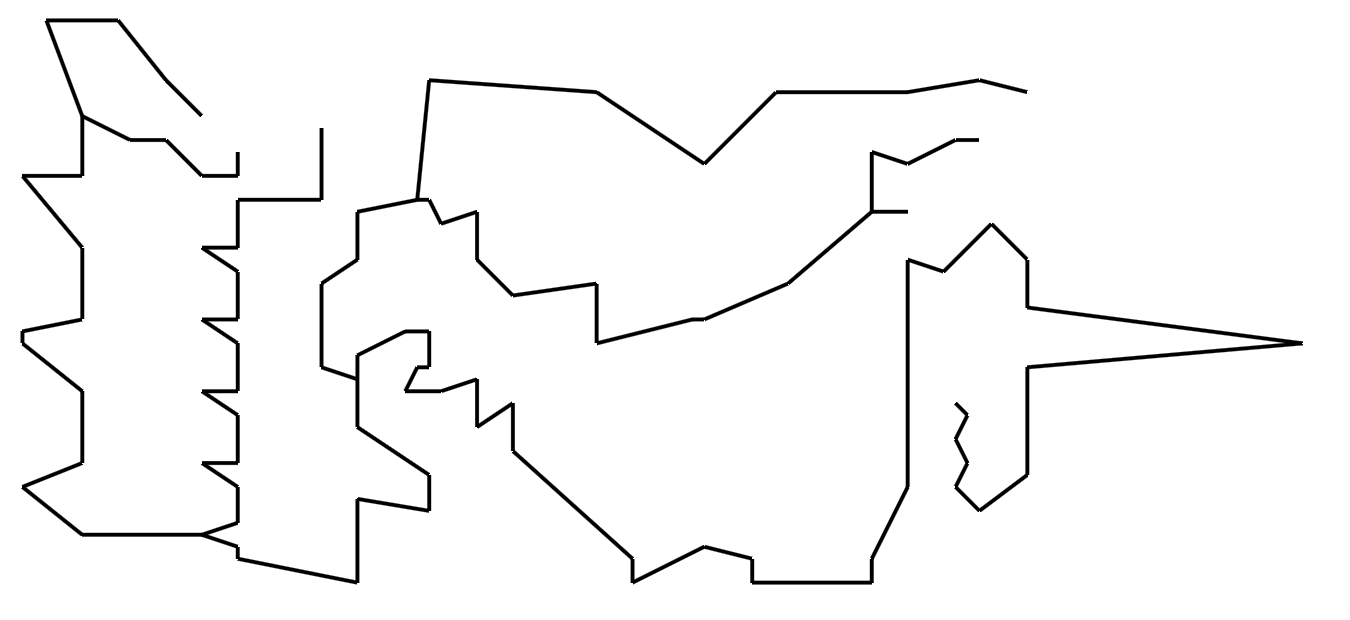}
  \caption{Splitting off}
\end{subfigure}%
\begin{subfigure}{.5\textwidth}
  \centering
  \includegraphics[width=.8\linewidth]{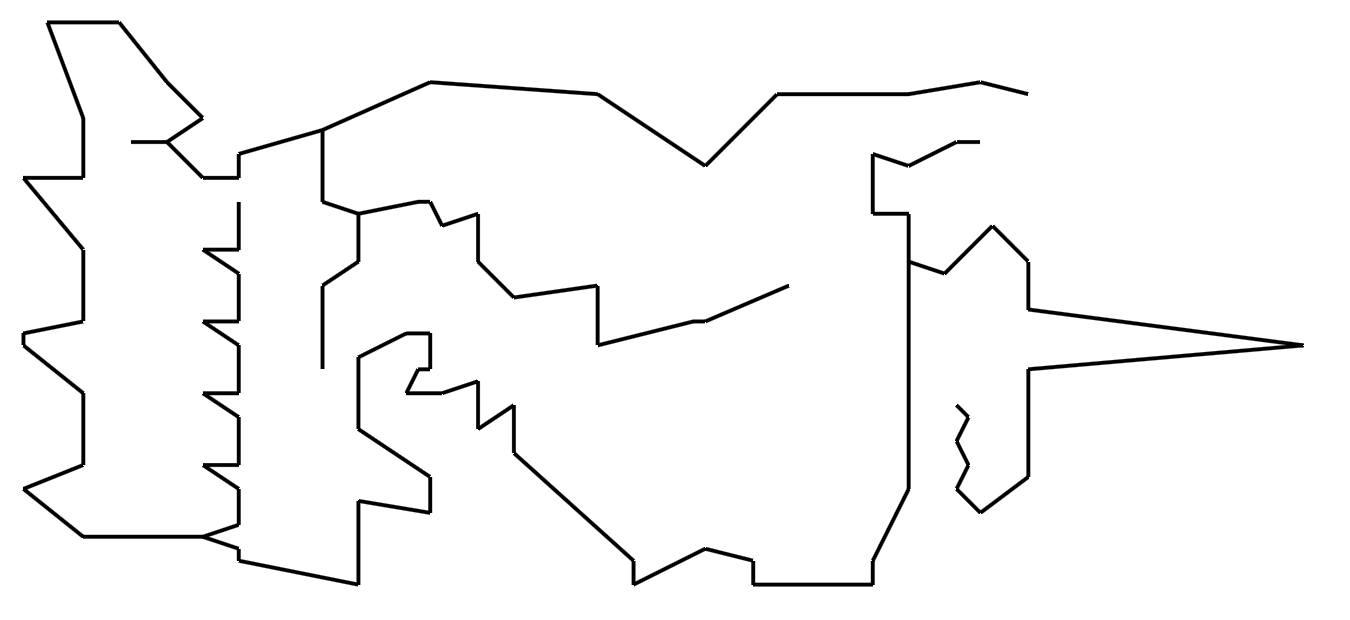}
  \caption{Splitting off + SwapRound}
\end{subfigure}
\caption{Sample trees from the various methods on VLSI instance XQF131 from Rohe \cite{Rohe}.}
\label{fig:fig}
\end{figure}

\begin{figure}
\begin{subfigure}{.5\textwidth}
  \centering
  \includegraphics[width=.8\linewidth]{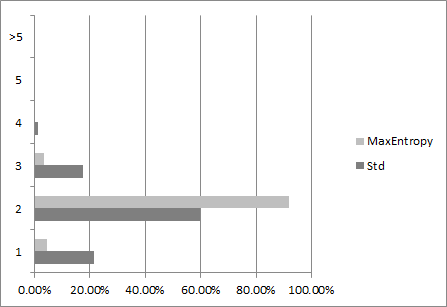}
  \caption{Euclidean TSPLIB}
\end{subfigure}%
\begin{subfigure}{.5\textwidth}
  \centering
  \includegraphics[width=.8\linewidth]{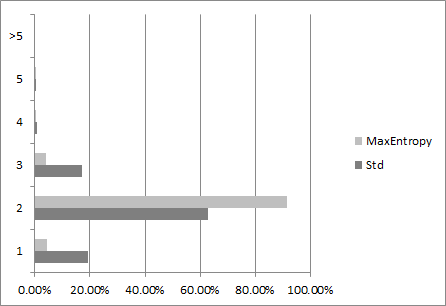}
  \caption{VLSI}
\end{subfigure}
\bigskip\bigskip
\begin{subfigure}{.5\textwidth}
  \centering
  \includegraphics[width=.8\linewidth]{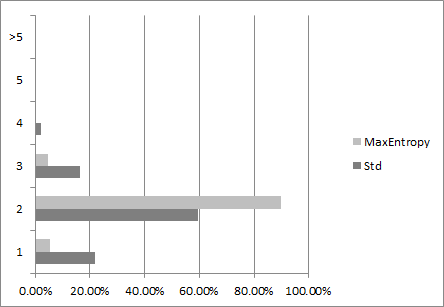}
  \caption{TSPLIB Non-Euclidean}
\end{subfigure}%
\begin{subfigure}{.5\textwidth}
  \centering
  \includegraphics[width=.8\linewidth]{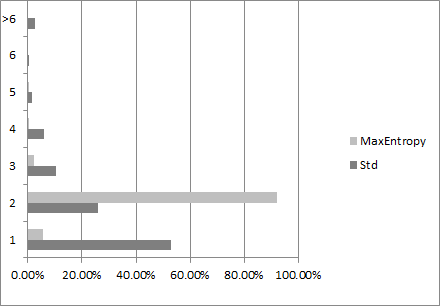}
  \caption{Graph}
\end{subfigure}
\caption{Degree distributions  of the minimum-cost spanning tree (used in standard Christofides) versus a tree sampled from the maximum entropy distribution.}
\label{fig:degree}
\end{figure}

One reason the matching costs are so much lower in the Best-of-Many Christofides' algorithm variants is that sampling a spanning tree from the subtour LP solution gives spanning trees such that a very high percentage of the vertices have degree two.  See Figure \ref{fig:fig} to compare the minimum-cost tree on the VLSI instance XQF131 with trees produced by the Best-of-Many Christofides' variants, and see Figure \ref{fig:degree} to see a comparison of the degrees of the minimum-cost spanning tree versus a tree sampled from the maximum entropy distribution for the four different instance types (trees from other variants had degree distributions similar to the trees from the maximum entropy distribution).  Thus the number of edges needed in the matching is much smaller.  These edges tend to be longer in the Best-of-Many Christofides' variants, because odd-degree vertices are rarer, and thus not as near to each other, but because the number of edges needed is much smaller, there is a significant reduction in the cost of the matching.  We summarize information about the matching costs in Table \ref{tab:matching}, where we give the average fraction of odd-degree nodes for the various algorithms, as well as the average cost of a matching edge (expressed in terms of percentage of the cost of an optimal tour).  For the TSPLIB and VLSI instances, 36 to 39\% of the min-cost spanning tree vertices are odd, whereas for the Best-of-Many variants, the number is between 8-12\%; however, the cost of each matching edge is roughly half to two-thirds more in these instances.  The graph instances are quite different; for these instances,  the min-cost spanning trees have nearly 66\% of vertices having odd degree, while the Best-of-Many variants have about the same percentages as before. For these instances, however, the cost of the matching edges are about the same for the standard Christofides'  algorithm and the Best-of-Many variants.

\begin{table}
{\footnotesize
\begin{tabular}{| l | r r | r r | r r | r r | r r | r r |}\hline
& \multicolumn{2}{|c|}{Std} & \multicolumn{2}{|c|}{ColGen} & \multicolumn{2}{|c|}{ColGen+SR} &\multicolumn{2}{|c|}{MaxEnt} & \multicolumn{2}{|c|}{Split} & \multicolumn{2}{|c|}{Split+SR} \\
&  \multicolumn{1}{|c}{Num} & \multicolumn{1}{c|}{Cost} & \multicolumn{1}{|c}{Num} & \multicolumn{1}{c|}{Cost} & \multicolumn{1}{|c}{Num} & \multicolumn{1}{c|}{Cost} & \multicolumn{1}{|c}{Num} & \multicolumn{1}{c|}{Cost} & \multicolumn{1}{|c}{Num} & \multicolumn{1}{c|}{Cost} & \multicolumn{1}{|c}{Num} & \multicolumn{1}{c|}{Cost}  \\ \hline
TSPLIB (E) & 39\%  & 0.89\% & 9.3\% & 1.3\% &  8.4\%  & 1.4\% & 8.0\% & 1.4\% & 7.6\% & 1.5\% & 7.8\% & 1.4\%  \\
VLSI &  36\% & 0.21\% & 12\% & 0.34\% & 11\% & 0.36\% & 8.6\% & 0.38\% & 8.3\% & 0.41\% & 8.3\% & 0.39\% \\
TSPLIB (N)  &  38\% & 0.48\% & 12\% & 0.63\% & 11\% & 0.67\% & 10\% & 0.66\% & 9.8\% & 0.69\% & 9.8\% & 0.66\% \\
Graph &  66\% & 1.9\% & 8.9\% & 1.7\% & 8.2\% & 1.7\% & 7.8\% & 1.7\% & 7.6\% & 1.5\% & 7.8\% & 1.5\% \\ \hline
\end{tabular}
}
\caption{Information about matchings for the various algorithms and instances.  `Num' indicates the average percentage of odd-degree vertices in the spanning tree, and `Cost' is the average cost of a matching edge expressed as a percentage of the cost of the optimal tour. }
\label{tab:matching}
\end{table}

Although we did not try to implement the various algorithms to be as fast as possible, it is instructive to compare their running times: see Table \ref{tab:running}.  Note that the reported running times do not include the time taken to solve the Subtour LP, although this was usually quite fast.  The standard Christofides' algorithm is quite fast compared to the other algorithms that must run Christofides' algorithm many times in addition to doing significant extra computational work.

\begin{table}
\begin{center}
\begin{tabular}{| l | r | r  | r  | r  | r | r |}\hline
	& \multicolumn{1}{|c|}{Std} &	\multicolumn{1}{|c|}{ColGen}	& \multicolumn{1}{|c|}{ColGen+SR}	& \multicolumn{1}{|c|}{MaxEnt}	& \multicolumn{1}{|c|}{Split}	 & \multicolumn{1}{|c|}{Split+SR} \\ \hline
TSPLIB (E) &	0.1	& 82.2 &	6900.1	& 1335.2 & 	3470.0 &	6671.7\\
VLSI	& 0.2 &	208.7 &	10414.4	& 27532.3 &	18128.2	& 29578.7 \\
TSPLIB (N) &	0.1 &	177.3	& 376.4	& 199.1	& 25.5 & 	48.5 \\
Graph	& 1.4 &	21.0 &	3401.1 & 	292.1 & 	2556.0	& 2821.6 \\\hline
\end{tabular}
\end{center}
\caption{Average running times (user time) for the variants on the different instance types. The running times do not take into account the time to compute the Subtour LP solution, although this was usually quite fast.}
\label{tab:running}
\end{table}

As a final observation, Schalekamp \cite{Schalekamp15} has pointed out that our degree distributions for the MSTs computed for Christofides' algorithm on Euclidean instances looks similar to the degree distributions for MSTs from random Euclidean instances.  In particular, suppose we draw points $n$ uniformly at random from the unit square, and compute the MST on these points.  Steele, Shepp, and Eddy \cite[Theorem 2]{SteeleSE87} have shown that as $n$ tends to infinity, the fraction of nodes of degree $k$ in the tree has a constant limit with probability 1.  In Table \ref{tab:euclidean-deg}, we compare the degree distribution from these random instances that Steele et al.\ obtained via simulation with the average degree of MSTs computed on the two sets of Euclidean instances.   For graph TSP instances, we can also compare degree distributions with the following model: suppose we compute an MST in a complete graph in which edge costs are drawn uniformly from [0,1].  Aldous \cite[Proposition 2]{Aldous90} gives a closed formula for the degree of a specified vertex in such a tree as $n$ tends to infinity.  In Table \ref{tab:graph-deg}, we compare the degree distribution from MSTs in this random model with the average degree of the MST computed in the graph TSP instances.  These appear to be less similar to the random instances; the fraction of degree one vertices is lower in the random instances, while the fraction of degree two and three vertices is higher.  Given the small number of graph TSP instances we use, it is hard to know if the results would be closer if we ran a larger number of instances, or whether there is some fundamental reason the two distributions are different.

\begin{table}
\begin{center}
\begin{tabular}{| l | r | r | r | r | r | r | r | } \hline
& 1 & 2 & 3 & 4 & 5 & 6 & $> 6$ \\ \hline
Random & 0.221 & 0.566 & 0.206 & 0.007 & 0.000 & 0.000 & 0.000 \\
TSPLIB (E) & 0.214 & 0.597 & 0.175 & 0.014 & 0.000 & 0.000 & 0.000 \\
VLSI & 0.192 & 0.627 & 0.171 & 0.170 & 0.010 & 0.000 & 0.000 \\ \hline
\end{tabular}
\end{center}
\caption{A comparison of the degree distribution from MSTs on random Euclidean instances (drawn from simulations performed by Steele, Shepp, and Eddy \cite[Table 1]{SteeleSE87}) and the degree distribution of MSTs computed on the Euclidean instances from TSPLIB and the VLSI instances.}
\label{tab:euclidean-deg}
\end{table}

\begin{table}
\begin{center}
\begin{tabular}{| l | r | r | r | r | r | r | r | } \hline
& 1 & 2 & 3 & 4 & 5 & 6 & $> 6$ \\ \hline
Random & 0.408 & 0.324 & 0.171 & 0.068 & 0.022 & 0.006 & 0.001 \\
Graph & 0.529 & 0.260 & 0.104 & 0.059 & 0.015 & 0.006 & 0.026 \\ \hline
\end{tabular}
\end{center}
\caption{A comparison of the degree distribution from MSTs on complete graphs with random weights (drawn from a computation of Aldous \cite[p.\ 396]{Aldous90}) and the degree distribution of MSTs computed on our graph TSP instances.}
\label{tab:graph-deg}
\end{table}

\section{Conclusions}
\label{sec:conc}

Our goal in this paper was to determine whether the empirical performance of the Best-of-Many Christofides' algorithms gives any reason to think they might be provably better than the Christofides' algorithm.  The answer to this question appears to be yes, with the large caveat that there are many heuristics for the traveling salesman problem (like Lin-Kernighan) with far better performance than Christofides' algorithm which have no provable performance guarantee at all.  We also wished to determine which variant might be most promising for further theoretical study.  For this question, it seems that the sampling methods have the most promise; that is, maximum entropy sampling or the SwapRound algorithm applied to some initial convex combination of trees.  However, because the good performance of these algorithms depends on taking the best result over a large number of samples drawn, one might have to argue that a good tour is produced with reasonable probability after multiple draws; it does not seem that one can argue that a good tour is produced in expectation from a single draw.  Because the average performance of these sampling methods does not seem significantly different from what happens when we construct an explicit convex combination (with column generation, or splitting off), it might be about as easy to prove that the expected value of a single draw of a sampling method is good as it is to prove that the average tour generated from an explicitly given convex combination is good, for a carefully constructed combination.

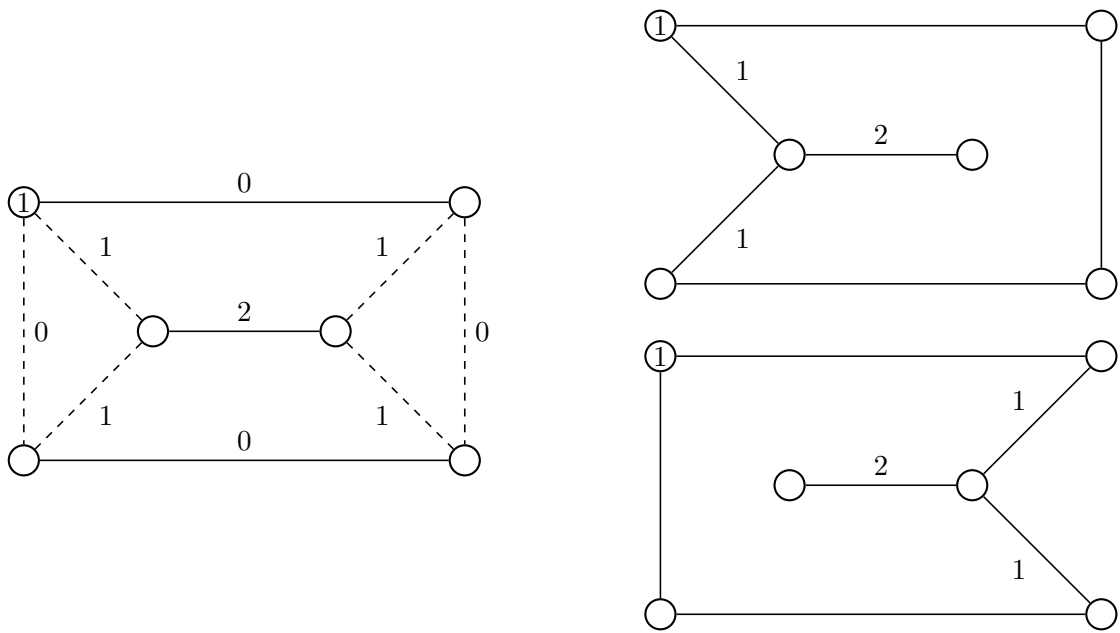
\begin{figure}
\begin{subfigure}{.5\textwidth}
\begin{center}
\begin{tikzpicture}
[auto, -,>=stealth',node distance=2cm,semithick,
vertex/.style={circle,draw=black,thick,inner sep=0pt,minimum size=4mm}]
\node[vertex] (a) {};
\node[vertex] (s) [above left=of a] {$1$};
\node[vertex] (b) [below left=of a] {} ;
\node[vertex] (c) [right=2 of a] {} ;
\node[vertex] (d) [above right=of c] {} ;
\node[vertex] (e) [below right=of c] {} ;
\path (s) edge[dashed] node {1} (a)
          edge[dashed] node {0} (b)
          edge node {0} (d)
      (a) edge[dashed] node {1} (b)
          edge node {2} (c)
      (b) edge node {0} (e)
      (c) edge[dashed] node {1} (d)
          edge[dashed] node[swap] {1} (e)
      (d) edge[dashed] node {0} (e);
\end{tikzpicture}
\end{center}
\end{subfigure}
\begin{subfigure}{.5\textwidth}
\begin{center}
\begin{tikzpicture}
[auto, -,>=stealth',node distance=2cm,semithick,
vertex/.style={circle,draw=black,thick,inner sep=0pt,minimum size=4mm}]
\node[vertex] (a) {};
\node[vertex] (s) [above left=of a] {$1$};
\node[vertex] (b) [below left=of a] {} ;
\node[vertex] (c) [right=2 of a] {} ;
\node[vertex] (d) [above right=of c] {} ;
\node[vertex] (e) [below right=of c] {} ;
\path (s) edge node {1} (a)
          edge (d)
      (a) edge node {1} (b)
          edge node {2} (c)
      (b) edge (e)
      (d) edge (e);
\end{tikzpicture}
\\[0.5cm]
\begin{tikzpicture}
[auto, -,>=stealth',node distance=2cm,semithick,
vertex/.style={circle,draw=black,thick,inner sep=0pt,minimum size=4mm}]
\node[vertex] (a) {};
\node[vertex] (s) [above left=of a] {$1$};
\node[vertex] (b) [below left=of a] {} ;
\node[vertex] (c) [right=2 of a] {} ;
\node[vertex] (d) [above right=of c] {} ;
\node[vertex] (e) [below right=of c] {} ;
\path (s) edge (b)
          edge (d)
      (a) edge node {2} (c)
      (b) edge (e)
      (c) edge node {1} (d)
          edge node[swap] {1} (e);
\end{tikzpicture}
\end{center}
\end{subfigure}

\caption{Example of Schalekamp and van Zuylen; the LP solution is on the left, and the convex combination of 1-trees is on the right, with each tree having weight 1/2.  In the LP solution,  dashed lines indicate LP value of 1/2, solid lines indicate LP value of 1.  Numbers represent the cost of the edges; other edge costs are implied by the triangle inequality.  The LP solution has a cost of 4, and any matching of the odd-degree vertices in the decomposition will have cost at least 2, so that the result of Christofides' algorithm will cost at least 3/2 times the cost of the LP solution.  One can verify that the LP solution is optimal via a duality argument.}
\label{fig:1tree}
\end{figure}

Very recently, Schalekamp and van Zuylen \cite{SchalekampvZ15} created an example that shows that for an {\em arbitrary} decomposition of an LP solution into a convex combination of 1-trees, it is not possible for the Best-of-Many Christofides' algorithm to do better than a $\frac{3}{2}$-approximation algorithm; we show the example in Figure \ref{fig:1tree}.   A {\em 1-tree} is a set of edges that has two edges incident to an arbitrary vertex, along with a spanning tree on the remaining vertices.  Held and Karp \cite{HeldK71} show that the Subtour LP solution can be decomposed into a convex combination of 1-trees.  This example provides further evidence that the randomization inherent in either maximum entropy sampling or SwapRound may well be needed to overcome worst-case decompositions.  Vygen's \cite{Vygen15} recent paper for the $s$-$t$ path TSP also uses the idea of modifying an initial decomposition of spanning trees so that the decomposition has specific properties that allow for a better performance guarantee than otherwise could be obtained.

In our limited experience, it seems that graph TSP is a significantly easier problem for the Best-of-Many variants, and this may bear further investigation both theoretically and empirically.  The quality of the solutions found by these algorithms relative to the standard Christofides' algorithm may indicate that these instances really are much easier than the general case of symmetric cost functions with triangle inequality.

\subsection*{Acknowledgments}

We thank Shayan Oveis Gharan for generously sharing his maximum entropy code with us; we used many of his implementation ideas in coding our own algorithm.  We thank Frans Schalekamp and Anke van Zuylen for allowing us to include their example of a bad case for Best-of-Many Christofides.  We thank Frans Schalekamp for several observations which he allowed us to use and include.

\bibliographystyle{abbrv}
\bibliography{bestofmany}

\end{document}